\documentclass[aps,prd,preprint,superscriptaddress,showpacs,floatfix,nofootinbib]{revtex4}

\usepackage{graphicx}
\def\ie{{\it i.e.}}
\def\eg{{\it e.g.}}

\def\nn{\nonumber}

\def\bwt{\begin{widetext}}
\def\ewt{\end{widetext}}
\def\be{\begin{equation}}
\def\ee{\end{equation}}
\def\bea{\begin{eqnarray}}
\def\eea{\end{eqnarray}}
\def\bean{\begin{eqnarray*}}
\def\eean{\end{eqnarray*}}
\def\bary{\begin{array}}
\def\eary{\end{array}}
\def\bit{\begin{itemize}}
\def\eit{\end{itemize}}
\def\GeV{\rm GeV}

\def\su5u1{SU(5) \times U(1)}
\def\fsu5u1{SU(5) \times U(1)'}
\def\so10{SO(10)}
\def\sq20{SO(10) \times SO(10)}

\begin{document}

\title{String Scale Gauge Coupling Unification with Vector-like
Exotics and Non-Canonical \boldmath{$U(1)_Y$} Normalization}

\author{V. Barger}
\affiliation{Department of Physics, University of Wisconsin, 
Madison, WI 53706, USA}

\author{Jing Jiang}
\affiliation{Institute of Theoretical Science, University of Oregon, 
Eugene, OR 97403, USA}

\author{Paul Langacker}
\affiliation{School of Natural Sciences, Institute for Advanced Study,
  Einstein Drive, Princeton, NJ 08540, USA}

\author{Tianjun Li}

\affiliation{George P. and Cynthia W. Mitchell Institute for
Fundamental Physics, Texas A$\&$M University, College Station, TX
77843, USA }

\affiliation{
Institute of Theoretical Physics, Chinese Academy of Sciences,
 Beijing 100080, China
}

\date{\today}

\begin{abstract}

We use a new approach to study string scale gauge coupling unification
systematically, allowing both the possibility of non-canonical
$U(1)_Y$ normalization and the existence of vector-like particles whose
quantum numbers are the same as those of the Standard Model (SM)
fermions and their Hermitian conjugates and the SM adjoint particles.
We first give all the independent sets (Yi) of particles that can be
employed to achieve $SU(3)_C$ and $SU(2)_L$ string scale gauge
coupling unification and calculate their masses. Second, for a
non-canonical $U(1)_Y$ normalization, we obtain string scale $SU(3)_C
\times SU(2)_L \times U(1)_Y$ gauge coupling unification by choosing
suitable $U(1)_Y$ normalizations for each of the Yi sets.
Alternatively, for the canonical $U(1)_Y$ normalization, we achieve
string scale gauge coupling unification by considering suitable
combinations of the Yi sets or by introducing additional independent
sets (Zi), that do not affect the $SU(3)_C \times SU(2)_L$
unification at tree level, and then choosing suitable combinations,
one from the Yi sets and one from the Zi sets.  We also briefly
discuss string scale gauge coupling unification in models with higher
Kac-Moody levels for $SU(2)_L$ or $SU(3)_C$.

\end{abstract}


\preprint{MADPH-05-1433, MIFP-06-35, OITS-769, hep-ph/0612206}

\maketitle

%\\[1ex]
%PACS: 11.25.Mj; 04.65.+e; 11.30.Pb; 12.60.Jv
%\\[1ex]
%Keywords: Grand Unified Theory; Symmetry Breaking; Extra Dimensions
\section{Introduction}

It is well known that the three gauge couplings in the Standard Model
(SM) do not unify for the canonical normalization of $U(1)_Y$ (Gauge
coupling unification in the SM can be realized via non-canonical
$U(1)_Y$ normalization~\cite{Barger:2005gn}). With supersymmetry (SUSY),
which provides an elegant solution to the gauge hierarchy problem,
gauge coupling unification can be approximately achieved in the
Minimal Supersymmetric Standard Model (MSSM) with unification
scale\footnote{The unification is not perfect: The $SU(3)$ and $SU(2)$
couplings unify at around $2 \times 10^{16}$ GeV, while the $SU(2)$
and $U(1)$ unification occurs around $3 \times 10^{16}$ GeV.}  $M_U$
around $2\times 10^{16}$
GeV~\cite{Langacker:1991an,Langacker:1992rq,Langacker:1995fk}. This
unification is based on two implicit assumptions: (1) the $U(1)_Y$
normalization is canonical; (2) there are no intermediate threshold
corrections.

However, the string scale $M_{\rm string}$ in weakly coupled heterotic 
string theory is~\cite{Dienes:1996du}
\begin{eqnarray}
M_{\rm string} = g_{\rm string} \times 5.3 \times 10^{17} ~{\rm
GeV}~,~\,
\label{eq:eq1}
\end{eqnarray}
where $g_{\rm string}$ is the string coupling constant. Since
$g_{\rm string} \sim {\cal O} (1)$, we have
\begin{eqnarray}
M_{\rm string} \approx 5 \times 10^{17} ~{\rm GeV}~.~\,
\end{eqnarray}
Thus, there exists a factor of approximately 20 to 25 between the MSSM
unification scale and the string scale.  In the strongly coupled
heterotic string theory or M-theory on $S^1/Z_2$~\cite{Horava:1995qa},
the eleven dimensional Planck scale can be the MSSM unification
scale~\cite{Witten:1996mz}.  However, our focus in this paper is on the
weakly coupled heterotic string theory.  The discrepancy between $M_U$
and $M_{\rm string}$ implies that the weakly coupled heterotic string
theory naively predicts the wrong values for the electroweak mixing
angle ($\sin^2\theta_W$) and strong coupling ($\alpha_3$) at the weak
scale. Because the weakly coupled heterotic string theory is one of
the leading candidates for a unified theory of the fundamental
particles and interactions in nature, how to achieve string scale
gauge coupling unification is an important question in string
phenomenology~\cite{Kaplunovsky:1987rp, Dixon:1990pc, Ibanez:1993bd,
Dienes:1995sv, Mayr:1995rx, Nilles:1995kb, Martin:1995wb,
Bachas:1995yt, Giedt:2002kb, Emmanuel-Costa:2005nh, Jiang:2006hf}.

In addition, there exist intermediate scales in many supersymmetric
theories, for example, the invisible axion models with an intermediate
Peccei-Quinn (PQ) symmetry breaking scale around $10^{11}$
GeV~\cite{PQ, review}, the see-saw neutrino models with intermediate
right-handed neutrino mass scale around $10^{14}$ GeV~\cite{Seesaw},
string models with gaugino condensation scale around $10^{13}$
GeV~\cite{Binetruy:1996xj}, or string constructions leading to new
vector-like matter not associated with any particular motivation.
Thus, there could exist intermediate threshold corrections. In
addition, there may exist threshold corrections close to the string
scale, for example, around $10^{16}$ GeV.  Similarly, heterotic
constructions often involve non-canonical $U(1)_Y$ embeddings (and
normalizations)~\cite{Dienes:1996du}. 

In this paper, assuming a low scale (TeV) supersymmetry, we
systematically study string scale gauge coupling unification by
introducing intermediate scale extra particles.  We introduce
vector-like particles whose quantum numbers are the same as those of
the SM fermions and their Hermitian conjugates, and SM adjoint
particles.  We do not consider particles which form complete (or
equivalent) $SU(5)$ multiplets because they do not change the relative
running among the gauge couplings at one-loop.
 
Our approach is different from previous approaches. We first list all
the independent sets Yi (defined in Section III) of particles that can
be employed to achieve $SU(3)_C$ and $SU(2)_L$ string scale gauge
coupling unification. Second, for the case of non-canonical $U(1)_Y$
normalization, we achieve string scale $SU(3)_C \times SU(2)_L
\times U(1)_Y$ unification by choosing suitable
$U(1)_Y$ normalizations.  Alternatively, for canonical $U(1)_Y$
normalization, string scale unification can be realized by considering
suitable combinations of Yi sets. We also introduce the independent
sets Zi (defined in subsection B in Section IV) of particles which do
not affect the relative running between the $SU(3)_C$ and $SU(2)_L$
gauge couplings at one-loop.  Then, string scale gauge coupling
unification can also be achieved by choosing suitable combinations of
one set of particles from Yi and one from Zi.  In quite a few cases
the masses for all the extra particles are roughly the same at the
intermediate scale of about $10^{15}$ GeV, or else one set is around
$10^{17}$ GeV, which can be considered as string scale threshold
corrections. In some cases there may exist extra particles with masses
around hundreds of GeV, which can be produced at the LHC or other
future colliders.  One can easily use our approach to discuss more
general and complicated cases of gauge coupling unification at the
string scale. Any set of additional particles that can be employed to
achieve the string scale gauge coupling unification can be decomposed
as a combination of the Yi and Zi sets of particles, plus complete
$SU(5)$ multiplets (these could merely have the quantum numbers of
$SU(5)$ complete multiplets; the full $SU(5)$ structure is
unnecessary).  With complete $SU(5)$ multiplet (or the equivalent), we
can shift the mass scales of the extra particles by splitting their
masses.  Furthermore, we briefly discuss string scale gauge coupling
unification in models with higher Kac-Moody levels for $SU(2)_L$ or
$SU(3)_C$.  A specific model with higher $SU(3)_C$ Kac-Moody level and
nocanonical $U(1)_Y$ normalization can be found
in~\cite{Emmanuel-Costa:2006ip}.

This paper is organized as follows: in Section II we present our
conventions and input data for the renormalization group equations
(RGEs).  We also list the SM vector-like and adjoint particles and
their contributions to the beta functions.  We study the string scale
$SU(3)_C$ and $SU(2)_L$ gauge coupling unification in Section III, and
the string scale $SU(3)_C$, $SU(2)_L$ and $U(1)_Y$ gauge coupling
unification in Section IV. In Section V, we briefly discuss string
scale gauge coupling unification in models with higher Kac-Moody
levels for $SU(2)_L$ or $SU(3)_C$. Discussion and conclusions are
presented in Section VI.

\section{RGEs and Extra Particles}

The relation between the string scale $M_{\rm string}$ and the string
coupling $g_{\rm string}$ is given in Eq.~(1). At the string scale,
the gauge couplings satisfy
\begin{eqnarray}
g_1 = g_2= g_3 = g_{\rm string}~,~\,
\label{eq:eq3}
\end{eqnarray}
where $g_1^2 \equiv k_Y g_Y^2$ with $k_Y=5/3$ for  
canonical normalization, and $g_Y$, $g_2$ and $g_3$ are
the gauge couplings for $U(1)_Y$, $SU(2)_L$ and $SU(3)_C$, respectively.
In addition, there exist threshold corrections to the
gauge coupling running in string models due to the massive string
states~\cite{Dienes:1996du} and in orbifold models due to the massive 
Kaluza-Klein states~\cite{Orbifold}.
Although these threshold corrections could be important in general, 
we will not consider them in this paper because we would like to 
give generic discussions which are model independent.

We define $\alpha_i=g_i^2/4\pi$ and denote the $Z$ boson
mass as $M_Z$.  The one-loop
renormalization group equations are given by
\begin{equation}
\frac{1}{\alpha_i(\mu)} = \frac{1}{\alpha_i(M_Z)} - \frac{b_i}{2 \pi}
\log \frac{\mu}{M_Z}\,,
\end{equation}
with $b \equiv (b_1, b_2, b_3) = (41/6k_Y,-19/6,-7)$ for the SM and $b
= (11/k_Y,1,-3)$ for the MSSM.  The two-loop RGE equations and
beta-functions can be found in the Appendix of \cite{Jiang:2006hf}.
For the numerical calculations, we use the central values of
$\alpha_3(M_Z) = 0.1189 \pm 0.0010$~\cite{Bethke:2006ac}, and
$\sin^2\theta_W (M_Z) = 0.23122\pm 0.00015$~\cite{Yao:2006px}.  For
simplicity, we usually assume a supersymmetry breaking scale of $300$
GeV.  This would be appropriate if all of the sparticles had that
value as a common mass.  However, in more realistic scenarios the mass
splittings, \eg, between squarks and sleptons lead to an effective
scale that is often much lower than the physical
masses~\cite{Langacker:1992rq,Langacker:1995fk}.  Hence, we will also
consider an effective scale of 50 GeV.

To achieve string scale gauge coupling unification, we only
introduce vector-like particles whose quantum numbers
are the same as those of the SM fermions and their Hermitian
conjugates, and SM adjoint particles. The 
 quantum numbers for additional particle multiplets
under the $SU(3)_C \times SU(2)_L \times U(1)_Y$ gauge symmetry 
and their contributions to the one-loop
 beta functions ($\Delta b$) are
\begin{eqnarray}
&& XQ + {\overline{XQ}} = {\mathbf{(3, 2, {1\over 6}) + ({\bar 3}, 2,
-{1\over 6})}}\,, \quad \Delta b =({1\over 5}, 3, 2)\,;\\ 
&& XU + {\overline{XU}} = {\mathbf{({\bar 3},  1, -{2\over 3}) + ({3},
1, {2\over 3})}}\,, \quad \Delta b = ({8\over 5}, 0, 1)\,;\\   
&& XD + {\overline{XD}} = {\mathbf{({\bar 3},  1, {1\over 3}) + ({3},
1, -{1\over 3})}}\,, \quad \Delta b = ({2\over 5}, 0, 1)\,;\\ 
&& XL + {\overline{XL}} = {\mathbf{(1,  2, {1\over 2}) + ({1},  2,
-{1\over 2})}}\,, \quad \Delta b = ({3\over 5}, 1, 0)\,;\\ 
&& XE + {\overline{XE}} = {\mathbf{({1},  1, {1}) + ({1},  1,
-{1})}}\,, \quad \Delta b = ({6\over 5}, 0, 0)\,;\\ 
&& XG = {\mathbf{({8}, 1, 0)}}\,, \quad \Delta b = (0, 0, 3)\,;\\ 
&& XW = {\mathbf{({1}, 3, 0)}}\,, \quad \Delta b = (0, 2, 0)\,.\,
\end{eqnarray}

Their two-loop beta functions are presented in \cite{Jiang:2006hf}.
We include two-loop running for the gauge couplings and one-loop
running for the Yukawa coupling in the RG evolution.  The Yukawa
couplings of the vector-like particles are not included. We do not
consider particles with the quantum number of complete $SU(5)$
multiplets because they do not change the relative running among the
gauge couplings at one-loop, although they can be employed to shift
the mass scales of the extra particles by splitting the masses of the
particles in the $SU(5)$ multiplets.

\section{\boldmath{$SU(3)_C \times SU(2)_L$} Gauge Coupling Unification}

In supersymmetric models with low scale supersymmetry breaking,
without introducing additional particles, the $SU(3)_C$ and $SU(2)_L$
gauge couplings unify at about $2 \times 10^{16}$ GeV.  To achieve 
$SU(3)_C \times SU(2)_L$ unification at the string
scale, we need to introduce sets of particles with $\Delta b_2 <
\Delta b_3$. The
independent sets of Yi particles, constructed from the extra particles
in Section II, that satisfy $\Delta b_2 < \Delta b_3$ are
\begin{eqnarray}
&& Y1: \quad XU + {\overline{XU}}\,, \quad \Delta b = (\frac{8}{5},0,1)\,; \\
&& Y2: \quad XD + {\overline{XD}}\,, \quad \Delta b = (\frac{2}{5},0,1)\,; \\
&& Y3: \quad XG+ XW\,, \quad \Delta b = (0,2,3)\,; \\
&& Y4: \quad XG + k (XQ+ {\overline{XQ}})+ l (XL+
{\overline{XL}})\,,\nn \\
&& \quad \quad \quad \Delta b = (\frac{k}{5}+\frac{3 l}{5}, 3 k + l, 2 k + 3) \,,
\end{eqnarray}
where $k=0, 1, 2$, and $l=0, 1, 2$ with $k + l \le 2$.  

The $SU(3)_C$ and $SU(2)_L$ gauge coupling unification at the string
scale can be realized by introducing any Yi set of particles, or any
combinations of the Yi sets.  Assuming that the new particles are
degenerate, their command mass can be determined from
Eqs.~(\ref{eq:eq1}) and (\ref{eq:eq3}).  For simplicity, we will not
study the cases with general combinations of the Yi sets here.  In the
cases Y1, Y2, Y3, and Y4 with $k+l=2$, we have $\Delta b_2 = \Delta
b_3-1$.  At one-loop level, the gauge coupling unification only
depends on the differences between the one-loop beta functions.
Hence, to achieve string scale $SU(3)_C \times SU(2)_L$ unification we
estimate that the mass scales ($\Lambda_Y$) at which these extra
particles are introduced are all approximately
\begin{eqnarray}
\Lambda_Y = 1.6 \times 10^{13}~\GeV\,. \quad 
\end{eqnarray}
For the other Y4 cases, the mass scales of the extra particles depend
on $k$ and $l$
\begin{eqnarray}
&& k = 0\,,~l = 0~:~\Lambda_Y = 1.6 \times 10^{16}~\GeV\,; \quad \nn \\
&&   k = 0\,,~l = 1~:~\Lambda_Y = 2.9 \times 10^{15}~\GeV\,; \quad \nn \\
&&  k = 1\,,~l = 0~:~\Lambda_Y = 2.9 \times 10^{15}~\GeV\,. \quad 
\end{eqnarray}
The particles in the set Y4 with $k = 0$ and $l = 0$ can be considered
as string scale threshold corrections.  For the two-loop predictions,
the actual values of the beta functions matter and these values are
shifted, as can be seen in the Table~\ref{tbl:a3}.

The mass scales of the extra particles can be shifted by introducing
complete $SU(5)$ multiplets but splitting their masses. For example,
consider the Y2 set $XD + {\overline{XD}}$. We can give them
vector-like mass $M_V$ with $M_V < 1.6\times10^{13}$, provide that we
also introduce vector-like particles $XL+{\overline{XL}}$ with mass
$M'_V \simeq M_V\times M_{\rm string}/(1.6\times10^{13}~{\rm GeV})$.
Above $M'_V$, we have complete ${\bf 5}+{\overline{\bf 5}}$
contributions to the gauge coupling RGE running, so that the $SU(3)_C
\times SU(2)_L$ gauge couplings can still be unified at the string
scale.  However, this approach introduces two mass scales
for the extra particles, so for simplicity we do not consider it
further.

\section{\boldmath{$SU(3)_C \times SU(2)_L\times U(1)_Y$} Gauge Coupling Unification}

At the string scale, where the $SU(3)_C$ and $SU(2)_L$ gauge couplings
unify, the $U(1)_Y$ gauge coupling in general does not coincide with
the other two couplings for a canonical $U(1)_Y$ normalization.  To
achieve string scale $SU(3)_C \times SU(2)_L\times U(1)_Y$
unification, the simplest possibilities are to consider suitable
non-canonical $U(1)_Y$ normalizations, add another Yi set at a
different scale, or introduce additional sets with $\Delta b_2 =
\Delta b_3$ at an intermediate scale.

\subsection{Non-Canonical \boldmath{$U(1)_Y$} Normalization }
\label{sec:non}

The $U(1)_Y$ normalization need not be canonical in string model
building \cite{Dienes:1996du,Blumenhagen:2005mu}, orbifold Grand
Unified Theories (GUTs) \cite{Orbifold,Li:2001tx} and their
deconstruction \cite{Arkani-Hamed:2001ca}, and in 4D GUTs with product
gauge groups. Similar to heterotic string models, we assume that $k_Y$
is a rational number.

\begin{figure}[htb]
\centering
\includegraphics[width=8cm]{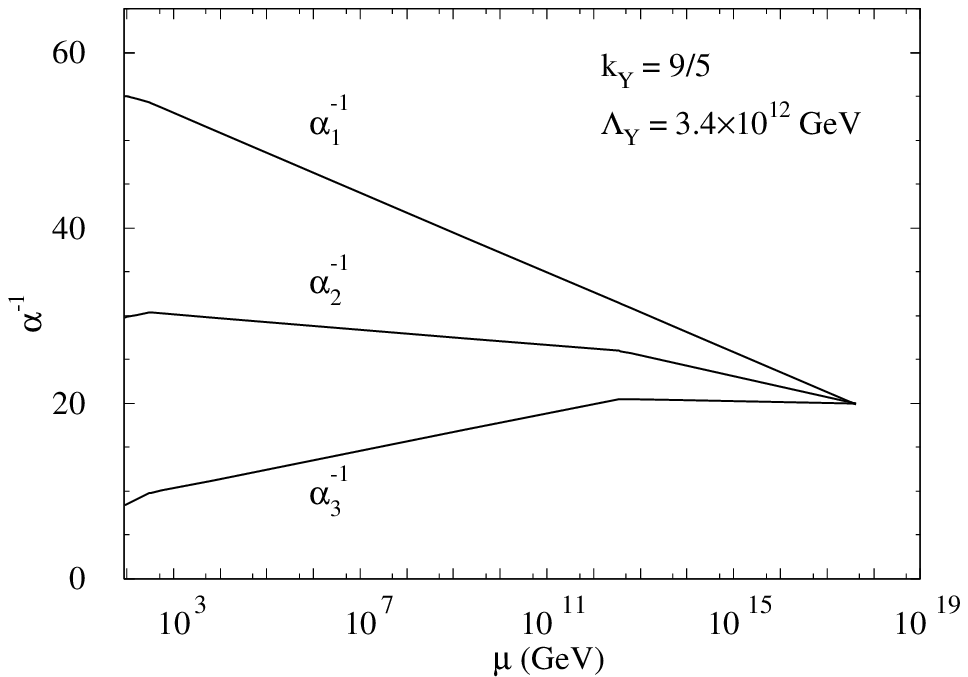}
\includegraphics[width=8cm]{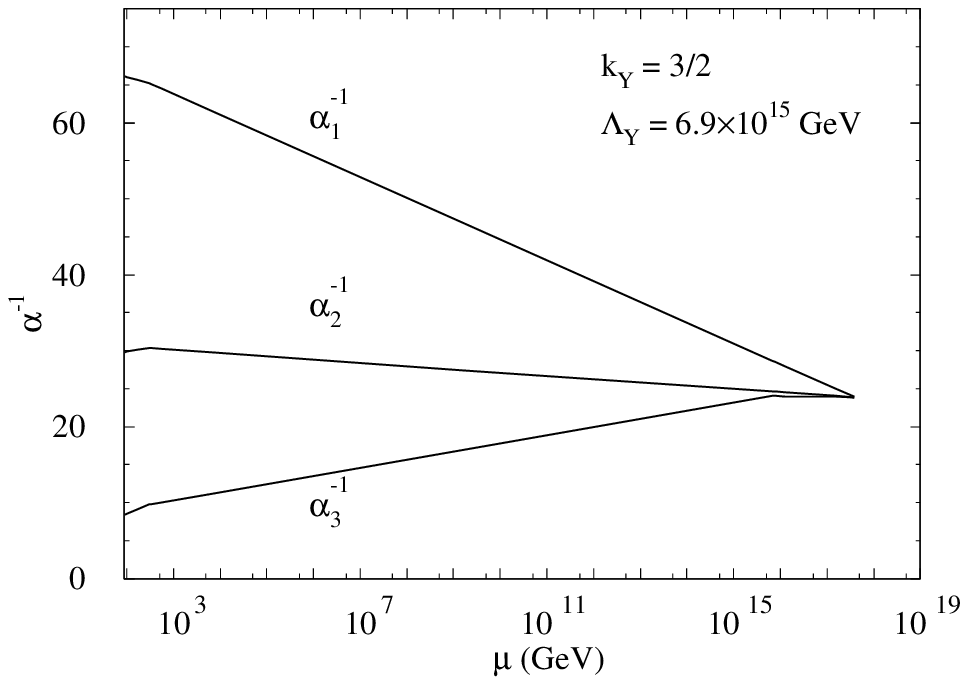}
\caption{Two-loop gauge coupling unification with non-canonical $U(1)_Y$ 
normalizations for the Yi sets.  Left: Y3 with $k_Y = 9/5$.  Right: Y4
(k=0, l=0) with $k_Y = 3/2$.}
\label{fig:non}
\end{figure}

Once the $SU(3)_C\times SU(2)_L$ gauge coupling unification at the
string scale is realized, we can unify the $U(1)_Y$ gauge coupling by
choosing a suitable $U(1)_Y$ normalization.  The non-canonical
$U(1)_Y$ normalizations required for the Yi sets of the particles are
given in Table~\ref{tbl:a3}.  The corresponding string scales $M_{\rm
string}$ can be obtained from Eq.~(1).  The two-loop prediction of the
scale at which the Yi particles are introduced is shown as
$\Lambda_Y$, together with gauge coupling $g_{\rm string}$ at the
string scale $M_{\rm string}$.  We also show the percentage deviation
of $\alpha_1^{-1}(M_{\rm string})$ from $\alpha_U^{-1}(M_{\rm
string})$, $\Delta=|\alpha_1^{-1}(M_{\rm string}) -
\alpha_U^{-1}(M_{\rm string})|/\alpha_U^{-1}(M_{\rm string})$, where
$\alpha_U^{-1}(M_{\rm string})$ is the unified gauge coupling for
$SU(3)_C \times SU(2)_L$ at the string scale.  The choice of $k_Y$ is
not unique.  We present the fractional number with the smallest
possible denominator, while requiring $\Delta$ defined above to be
less than $5\%$.  In Fig.~\ref{fig:non} we show the two-loop gauge
couplings for the cases Y3 with $k_Y = 9/5$, and Y4 ($k=0,\ l=0$) with
$k_Y = 3/2$.  It is interesting to point out that although the
canonical gauge coupling unification can be realized at one-loop level
for the Y3 set of particles~\cite{Bachas:1995yt}, this is no longer
true at two-loop level.
\begin{table}[htb]
\begin{center}
\begin{tabular}{|c|c|c|c|c|c|}
\hline
Y's & $\Lambda_Y$  & $g_{\rm string}$ & $k_Y$    & $k_Y$/(5/3) &
$\Delta$ (\%)\\  
\hline
Y1            &  $1.8 \times 10^{12}$  & 0.725 & 9/7 & 0.771  & 3.1
\\ 
Y2            &  $1.8 \times 10^{12}$  & 0.726 & 29/20 & 0.870  & 2.1 \\
Y3            &  $3.4 \times 10^{12}$  & 0.794 & 9/5 & 1.080   & 3.2 \\   
Y4 (k=0, l=0) &  $6.9 \times 10^{15}$  & 0.725 & 3/2 & 0.900 
& 0.9  \\  
Y4 (k=0, l=1) &  $1.0 \times 10^{15}$  & 0.741 & 29/19 & 0.916  & 1.7   \\  
\hline
Y4 (k=0, l=2) &  $4.7 \times 10^{12}$  & 0.791 & 8/5 & 0.960    & 2.3  \\  
Y4$^*$ (k=0, l=2) &  $4.3 \times 10^{12}$  & 0.818 & 18/11 & 0.982    & 0.0 \\
\hline
Y4 (k=1, l=0) &  $9.9 \times 10^{14}$  &  0.776 & 17/10 & 1.020  & 2.1  \\
Y4$^*$ (k=1, l=0) &  $9.3 \times 10^{14}$  &  0.803 & 7/4 & 1.050  & 0.7  \\
\hline
Y4 (k=1, l=1) &  $3.8 \times 10^{12}$  & 0.887 & 31/15 & 1.240   & 2.2  \\  
Y4 (k=2, l=0) &  $3.0 \times 10^{12}$  & 1.051 & 3 & 1.800   & 3.2  \\  
\hline
\end{tabular}
\end{center}
\caption{The required mass scales (for $SU(3)_C \times SU(2)_L$
unification) and $U(1)_Y$ normalization of the Yi sets (for $U(1)_Y$
unification).  The rows with an asterisk are for an effective SUSY
breaking scale of 50 GeV.}
\label{tbl:a3}
\end{table}

The above Yi sets can be categorized according to their $k_Y$ values,
depending on whether $k_Y$ is smaller than (AY) or greater than (BY) $5/3$:
\begin{eqnarray}
&& {\rm Case~AY}: ~Y1, ~Y2, ~Y4~(k=0,~l=0), ~Y4~(k=0,~l=1), ~Y4~(k=0,~l=2)\,;\nn \\ 
&& {\rm Case~BY}: ~Y3, ~Y4 ~(k=1,~l=0), ~Y4 ~(k=1,~l=1), ~Y4 ~(k=2,~l=0)\,.\nn
\end{eqnarray}

Using the effective SUSY scale of 50 GeV instead of 300 GeV could
potentially change the $k_Y$ value and $\Lambda_Y$ scale for each
case.  In Table~\ref{tbl:a3}, we show the effect of using 50 GeV for
two cases with $k_Y$ close to $5/3$, Y4 ($k=0,\ l=2$) and Y4 ($k=1,\
l=0$).  We see from these examples that the change is less than $10\%$
for $\Lambda_Y$ and even smaller for $k_Y$.

\subsection{Canonical \boldmath{$U(1)_Y$} Normalization }

Two ways to achieve string scale $SU(3)_C \times SU(2)_L\times U(1)_Y$
gauge coupling unification for canonical $U(1)_Y$ normalization are by
combining Case AY and Case BY sets, or by
introducing another set of particles with $\Delta b_2 =
\Delta b_3$ at an intermediate scale.

\begin{figure}[htb]
\centering
\includegraphics[width=8cm]{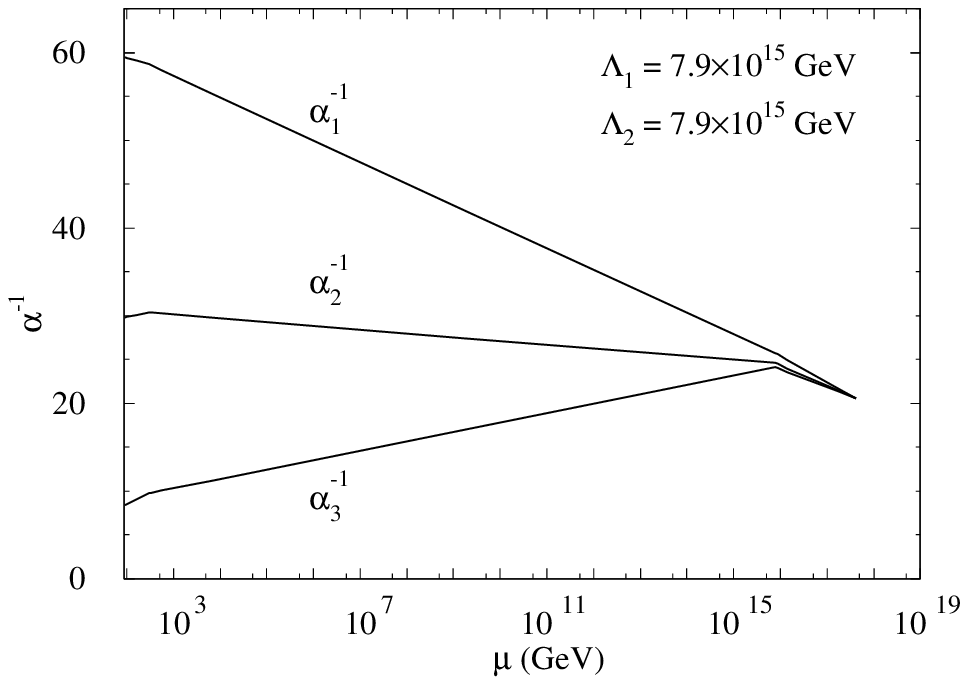}
\includegraphics[width=8cm]{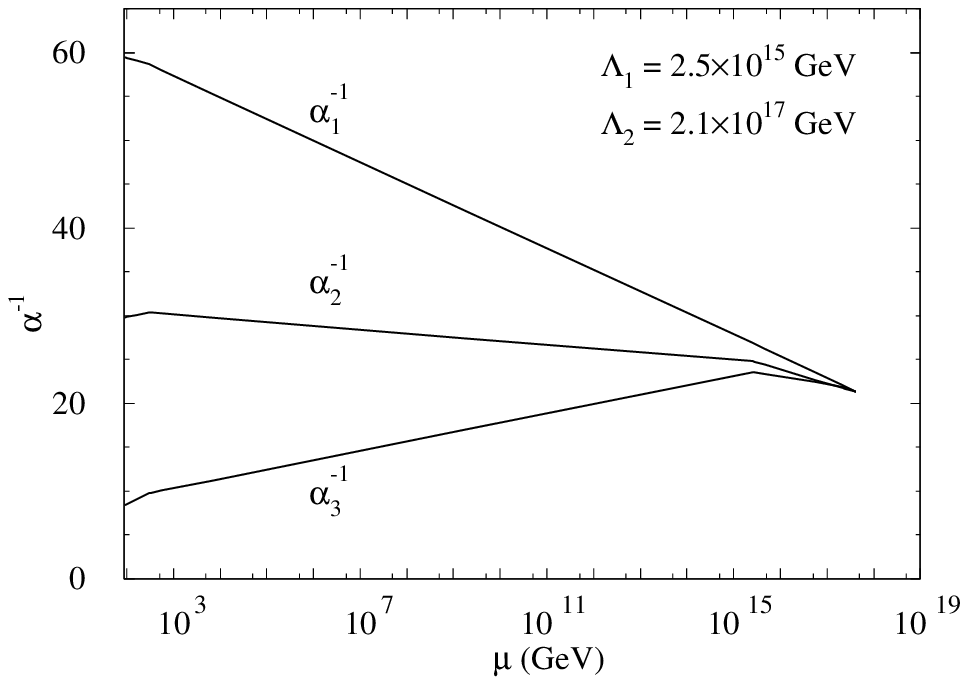}
\caption{Two-loop gauge coupling unification for one set of
particles from Case AY and one from BY.  Left: Y4 ($k=0,\ l=1$) and Y4
($k=1,\ l=1$) at $7.9 \times 10^{15}$ GeV.  Right: Y4 ($k=1,\ l=0$) at $2.5
\times 10^{15}$ GeV and Y4 ($k=0,\ l=0$) at $2.1 \times 10^{17}$ GeV.}
\label{fig:AandC}
\end{figure}

(1) If we introduce one set of particles at mass scale $\Lambda_1$
from the AY sets and another set at scale $\Lambda_2$ from BY, we are
able to realize gauge coupling unification at the string scale, by
adjusting $\Lambda_1$ and $\Lambda_2$ to satisfy Eqs.~(\ref{eq:eq1})
and (\ref{eq:eq3}). In Table \ref{tbl:AandC}, we present the mass
scales in GeV and the corresponding unified gauge couplings $g_{\rm
string}$. For the cases Y4 ($k=0,\ l=1$) with Y4 ($k=1,\ l=1$); and Y4
($k=0,\ l=2$) with Y4 ($k=1,\ l=0$), we have $\Lambda_1
\sim \Lambda_2$, \ie, almost a common mass scale for
all the extra particles. For the cases Y1 with Y4 ($k=1,\ l=0$); Y4
($k=0,\ l=0$) with Y4 ($k=1,\ l=0$); and Y4 ($k=0,\ l=2$) with Y4
($k=2,\ l=0$), one set of the extra particles could be considered as
string scale threshold correction because $\Lambda \sim 10^{17}$ GeV.
We show the couplings for the cases Y4 ($k=0,\ l=1$) with Y4 ($k=1,\
l=1$), and Y4 ($k=0,\ l=2$) with Y4 ($k=2,\ l=0$) in
Fig. \ref{fig:AandC}.  For the case Y4 ($k=0,\ l=1$) with Y4 ($k=1,\
l=0$), and the cases Y4 ($k=0,\ l=0$) with Y3, Y4 ($k=1,\ l=1$), or Y4
($k=2,\ l=0$), we cannot achieve string scale gauge coupling
unification.

\begin{table}[htb]
\begin{center}
\begin{tabular}{|c|ccc|ccc|}
\hline
         & \multicolumn{3}{c|}{Y3} & \multicolumn{3}{c|}{Y4 (k=1, l=0)} \\ 
\hline
             & $\Lambda_{Y3}$ & $\Lambda_Y$ & $g_{\rm string}$ &
$\Lambda_{Y4(1,0)}$ & $\Lambda_Y$ & $g_{\rm string}$  \\
\hline
Y1           & $4.1 \times 10^{13}$ & $2.9 \times 10^{16}$ & 0.777 & $1.5
\times 10^{15}$ & $1.6 \times 10^{17}$ & 0.772 \\ 
Y2           & $1.5 \times 10^{14}$ & $7.0 \times 10^{15}$ & 0.769 &  $2.1
\times 10^{15}$ & $8.8 \times 10^{16}$ & 0.769 \\
Y4 (k=0, l=0) & - & - & -   &
$2.5 \times 10^{15}$ & $2.1 \times 10^{17}$ & 0.768  \\
Y4 (k=0, l=1) & $5.6 \times 10^{14}$ & $2.9 \times 10^{16}$ & 0.769 &  
- & - & - \\
Y4 (k=0, l=2) & $6.3 \times 10^{15}$ & $2.5 \times 10^{14}$ & 0.792 &  
$7.8 \times 10^{15}$ & $7.9 \times 10^{15}$ & 0.781\\
\hline
         & \multicolumn{3}{c|}{Y4 (k=1, l=1)} & \multicolumn{3}{c|}{Y4
(k=2, l=0)} \\ 
\hline
             & $\Lambda_{Y4(1,1)}$ & $\Lambda_Y$ & $g_{\rm string}$ &
$\Lambda_{Y4(2,0)}$ & $\Lambda_Y$ & $g_{\rm string}$  \\
\hline
Y1           & $4.1 \times 10^{14}$ & $2.8 \times 10^{15}$ & 0.809 &
$4.0 \times 10^{15}$ & $2.2 \times 10^{14}$ & 0.810 \\ 
Y2           & $2.1 \times 10^{15}$ & $4.7 \times 10^{14}$ & 0.786 &
$1.7 \times 10^{16}$ & $4.7 \times 10^{13}$ &  0.779 \\
Y4 (k=0,\ l=0) & - & - & - &
- & - & -  \\
Y4 (k=0,\ l=1) & $7.9 \times 10^{15}$ & $7.9 \times 10^{15}$ & 0.781 &  
$4.6 \times 10^{16}$ & $3.1 \times 10^{15}$ &  0.776 \\
Y4 (k=0,\ l=2) & $5.0 \times 10^{16}$ & $3.6 \times 10^{13}$ & 0.806 &  
$1.5 \times 10^{17}$ & $1.3 \times 10^{13}$ &  0.806 \\
\hline
\end{tabular}
\end{center}
\caption{The mass scales in GeV for the particles in the Yi sets and the 
corresponding unified gauge couplings  $g_{\rm string}$.}
\label{tbl:AandC}
\end{table}

One could, of course, consider more complicated cases involving more
sets of particles.

(2) As another way to achieve $SU(3)_C \times SU(2)_L\times U(1)_Y$
unification, we introduce the following particle sets Zi 
with $\Delta b_2 = \Delta b_3$ at another scale:
\begin{eqnarray}
&&Z1: XE + {\overline{XE}}\,,~\Delta b = ({6\over 5}, 0,
0)\,; \\
&&Z2: XQ+ {\overline{XQ}}+XU + {\overline{XU}}\,, ~\Delta
b =({9\over 5}, 3, 3)\,; \\
&&Z3: XQ+ {\overline{XQ}}+XD + {\overline{XD}}\,, ~\Delta
b =({3\over 5}, 3, 3)\,; \\
&&Z4: XL+ {\overline{XL}}+XU + {\overline{XU}}\,, ~\Delta
b =({11\over 5}, 1, 1)\,; \\
&&Z5: XG+XW+XL+ {\overline{XL}}\,, ~\Delta b =({3\over 5},
3, 3)\,; \\
&&Z6: XG+XW+XQ+ {\overline{XQ}}\,, ~\Delta b =({1\over5},
5, 5)\,; \\
&&Z7: XG+ n(XQ+ {\overline{XQ}})+(3-n)(XL+{\overline{XL}})\,, \nn \\
&&\quad\quad \Delta b =({9\over 5}-{2\over 5}n, 3+2n, 3+2n)\,; \\
&&Z8: XW+ m (XU+ {\overline{XU}})+(2-m)(XD+{\overline{XD}})\,, \nn \\
&&\quad\quad \Delta b = ({4\over 5}+{6\over 5}m, 2, 2)\,,
\end{eqnarray}
where $m=0, 1, 2$, and $n=0, 1, 2, 3$.  The Z3 and Z5 sets of
particles give the same contribution to $\Delta b$, so we only show
the results for the $Z3$ set\footnote{If we only introduce one Zi set
and no Yi sets, gauge coupling unification can be achieved at about
$2\times 10^{16}$ GeV at one-loop level by considering non-canonical
$U(1)_Y$ normalizations.  $k_Y > 5/3$ for the sets Z2, Z3, Z5, Z6, Z7,
Z8 (m=0) and $k_Y < 5/3$ for Z1, Z4, Z8 (m=2).}.  The Z8 ($m = 1$) set
satisfies $\Delta b_1=\Delta b_2 = \Delta b_3$, so we do not consider
it here.  The sets $Z1 + Z2$ and $Z3 + Z4 - Z2$ form complete $SU(5)$
multiplets and would not contribute if they are degenerate.

We can introduce one combination of Yi sets and another from the Zi.
For simplicity, we only consider the cases with one set from Yi and
one from Zi. The $SU(3)_C\times SU(2)_L \times U(1)_Y$ unification can
be achieved by the following combinations:
\begin{eqnarray}
&& {\rm Case~AYZ}: ~Y1, ~Y2, ~Y4~(k=0,~l=0,1,2), ~Y4~(k=1,l=0)~ \nn \\
&&\quad \quad \quad {\rm with}~ Z2, ~Z3,~Z5,~Z6,~Z7,~Z8~(m=0)\,;\nn \\ 
%% && {\rm Case~BYZ}: ~Y3 ~{\rm with}~ Z8~(m = 1)\,; \nn \\
&& {\rm Case~BYZ}: ~Y3, ~Y4 ~(k=1,~l=1),~Y4~(k=2,~l=0) ~{\rm with}~ Z1, ~Z4,
~Z8~(m=2)\,. \nn 
\end{eqnarray}
In all cases, the Yi sets guarantee the unification of $SU(3)_C$ and
$SU(2)_L$ at the string scale, and the Zi sets of particles ensure the
unification of the $U(1)_Y$ gauge coupling.

\begin{figure}[htb]
\centering
\includegraphics[width=8cm]{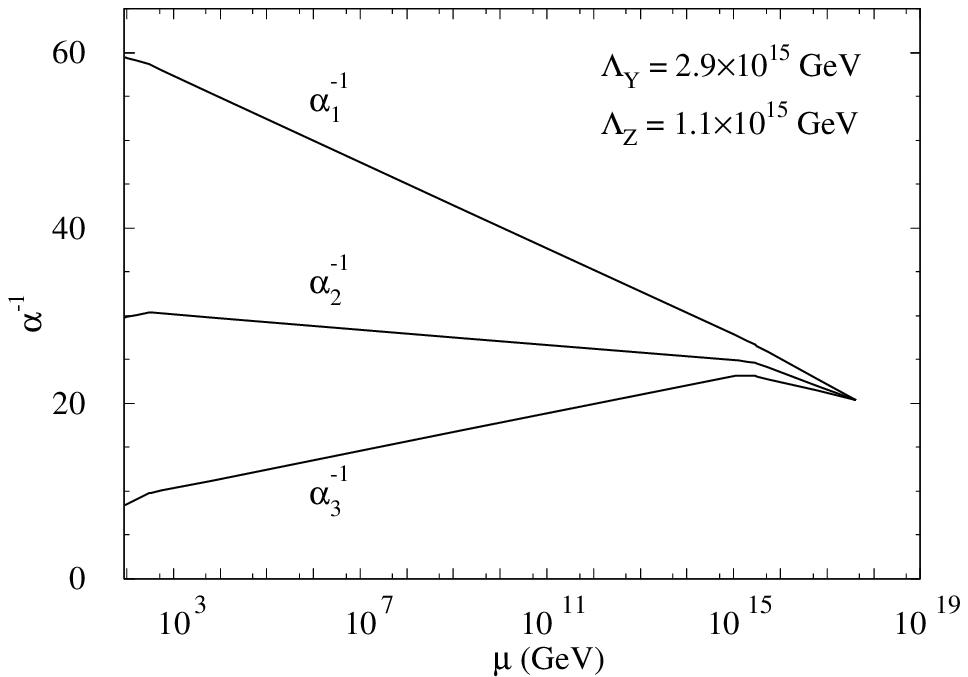}
\includegraphics[width=8cm]{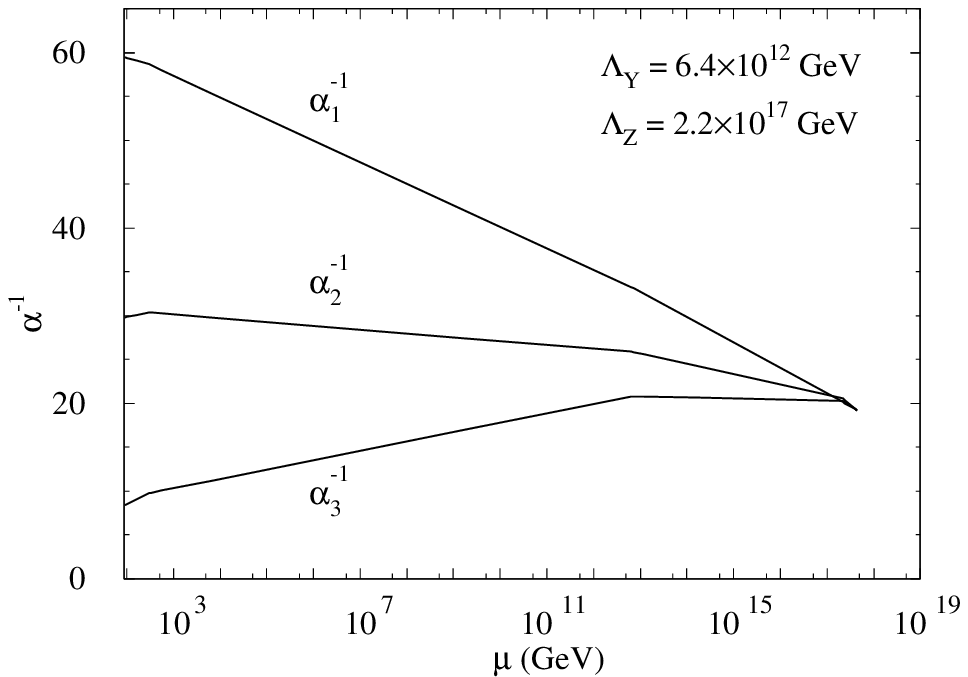}
\caption{Two-loop gauge coupling unification for Cases AYZ.  Left: Y4
($k=0,\ l=1$) at $2.9 \times 10^{15}$ GeV and Z3 at $1.1 \times
10^{15}$ GeV.  Right: Y4 (k=0, l=2) at $6.4 \times 10^{12}$ GeV and Z7
(n=3) at $2.2 \times 10^{17}$ GeV.}
\label{fig:caseAYZ}
\end{figure}

We denote $\Lambda_Y$ and $\Lambda_Z$ as the mass scales for the Yi
and Zi sets of particles, respectively.  The energy scales $\Lambda_Y$
and $\Lambda_Z$ as well as the corresponding unified gauge couplings
$g_{\rm string}$ for Case AYZ are listed in Table~\ref{tbl:AAU}.
There are several cases with $\Lambda_Y \sim
\Lambda_Z$.  There are also several cases with
$\Lambda_Z \sim 10^{17}$ GeV,which can be considered as string scale
threshold corrections.  In Fig.~\ref{fig:caseAYZ} we plot the two-loop
gauge coupling unification for Case AYZ, Y4($k=0,\
l=0$) with Z7 ($n=1$) as an example of the case with $\Lambda_Y \sim
\Lambda_Z$, and Y4 ($k=0,\
l=2$) with Z7 ($n=3$) as an example of the case with $\Lambda_Z \sim
10^{17}$.
\begin{table}[htb]
\begin{center}
\begin{tabular}{|c|ccc|ccc|}
\hline
         & \multicolumn{3}{c|}{Y1} & \multicolumn{3}{c|}{Y2} \\ 
\hline
             & $\Lambda_Z$ & $\Lambda_Y$ & $g_{\rm string}$ &
$\Lambda_Z$ & $\Lambda_Y$ & $g_{\rm string}$  \\
\hline
Z2       & $2.0 \times 10^{5}$  & $9.2 \times 10^{11}$ & 1.213  & $6.2
\times 10^{10}$ & $1.3 \times 10^{12}$ & 0.893  \\ 
Z3       & $2.8 \times 10^{11}$ & $1.4 \times 10^{12}$ & 0.870  & $1.5
\times 10^{14}$ & $1.5 \times 10^{12}$ & 0.796 \\ 
Z6       & $4.0 \times 10^{14}$ & $2.3 \times 10^{12}$ & 0.836  & $7.8
\times 10^{15}$ & $2.0 \times 10^{12}$ & 0.782 \\ 
Z7 (n=1) & $4.7 \times 10^{13}$ & $3.5 \times 10^{12}$ & 0.882  & $2.1
\times 10^{15}$ & $2.5 \times 10^{12}$ & 0.804 \\ 
Z7 (n=2) & $1.6 \times 10^{15}$ & $2.3 \times 10^{12}$ & 0.853  & $1.7
\times 10^{16}$ & $2.0 \times 10^{12}$ & 0.790 \\ 
Z7 (n=3) & $7.9 \times 10^{15}$ & $1.9 \times 10^{12}$ & 0.841  & $4.3
\times 10^{16}$ & $1.8 \times 10^{12}$ & 0.784  \\ 
Z8 (m=0) & $1.8 \times 10^{5}$  & $1.6 \times 10^{12}$ & 0.946  & $5.1
\times 10^{10}$ & $1.4 \times 10^{12}$ & 0.824 \\ 
\hline
         & \multicolumn{3}{c|}{Y4 (k=0, l=0)} & \multicolumn{3}{c|}{Y4
(k=0, l=1)} \\ 
\hline
             & $\Lambda_Z$ & $\Lambda_Y$ & $g_{\rm string}$ &
$\Lambda_Z$ & $\Lambda_Y$ & $g_{\rm string}$  \\
\hline
Z2       & $3.6 \times 10^{12}$  & $7.3 \times 10^{15}$ & 0.837  &
$2.3 \times 10^{13}$ & $1.1 \times 10^{15}$ & 0.838 \\ 
Z3       & $1.1 \times 10^{15}$  & $7.0 \times 10^{15}$ & 0.775  &
$1.1 \times 10^{15}$ & $2.9 \times 10^{15}$ & 0.785 \\ 
Z6       & $2.1 \times 10^{16}$  & $7.4 \times 10^{15}$ & 0.766  &
$3.4 \times 10^{16}$ & $1.1 \times 10^{15}$ & 0.777 \\ 
Z7 (n=1) & $7.9 \times 10^{15}$  & $7.9 \times 10^{15}$ & 0.781  &
$1.5 \times 10^{16}$ & $1.2 \times 10^{15}$ & 0.790  \\ 
Z7 (n=2) & $3.8 \times 10^{16}$  & $7.4 \times 10^{15}$ & 0.771  &
$5.6 \times 10^{16}$ & $1.1 \times 10^{15}$ & 0.781  \\ 
Z7 (n=3) & $7.5 \times 10^{16}$  & $7.2 \times 10^{15}$ & 0.767  &
$9.9 \times 10^{16}$ & $1.0 \times 10^{15}$ & 0.778 \\ 
Z8 (m=0) & $3.2 \times 10^{12}$  & $6.9 \times 10^{15}$ & 0.794  &
$2.1 \times 10^{13}$ & $1.0 \times 10^{15}$ & 0.801 \\ 
\hline
         & \multicolumn{3}{c|}{Y4 (k=0, l=2)} & \multicolumn{3}{c|}{} \\ 
\hline
             & $\Lambda_Z$ & $\Lambda_Y$ & $g_{\rm string}$ &
 & & \\
\hline
Z2       & $3.9 \times 10^{15}$ & $6.0 \times 10^{12}$ & 0.839 &  &  &  \\ 
Z3       & $3.9 \times 10^{16}$ & $6.1 \times 10^{12}$ & 0.812 &  &  &  \\ 
Z6       & $1.3 \times 10^{17}$ & $6.6 \times 10^{12}$ & 0.808 &  &  &  \\ 
Z7 (n=1) & $8.4 \times 10^{16}$ & $7.0 \times 10^{12}$ & 0.815 &  &  &  \\ 
Z7 (n=2) & $1.6 \times 10^{17}$ & $6.6 \times 10^{12}$ & 0.810 &  &  &  \\ 
Z7 (n=3) & $2.2 \times 10^{17}$ & $6.4 \times 10^{12}$ & 0.809 &  &  &  \\ 
Z8 (m=0) & $3.8 \times 10^{15}$ & $5.9 \times 10^{12}$ & 0.821 &  &  &  \\ 
\hline
\end{tabular}
\end{center}
\caption{The mass scales in GeV for the particles in the Yi and Zi
sets and the corresponding unified gauge couplings $g_{\rm string}$
for case AYZ.}
\label{tbl:AAU}
\end{table}

We show the mass scales and the corresponding unified gauge couplings
for Case BYZ in Table~\ref{tbl:CCU}. For the cases Y4 ($k=2,\ l=0$)
with Z4, we find that $\Lambda_Z < M_Z$ and $\Lambda_Y \sim 10^{13}$
GeV.  The low $\Lambda_Z$ value is very sensitive to the supersymmetry
and the string scale threshold corrections, and should be considered
an order of magnitude estimate only.  In particular, for a higher
effective supersymmetric mass scale, one can raise the $\Lambda_Z$
above 200 GeV, which is the Tevatron bound~\cite{Affolder:1999bs}.
Then, Z4 set of particles may be observable at the
LHC.  In practice, it is difficult to obtain a higher effective scale
in the SUSY breaking schemes with physical masses not too much higher
than the TeV scale.  That is because the effective scale is very
sensitive to the mass splittings, and those schemes (such as most
SUGRA and gauge mediated models) in which the colored sparticles are
typically heavier than the uncolored ones tend to give a low effective
mass~\cite{Langacker:1992rq,Langacker:1995fk}.

\begin{table}[htb]
\begin{center}
\begin{tabular}{|c|ccc|ccc|}
\hline
         & \multicolumn{3}{c|}{Y3} & \multicolumn{3}{c|}{Y4 (k=1, l=0)} \\ 
\hline
         & $\Lambda_Z$ & $\Lambda_Y$ & $g_{\rm string}$ & $\Lambda_Z$
& $\Lambda_Y$ & $g_{\rm string}$  \\
\hline
Z1       & $8.3 \times 10^{14}$  & $6.4 \times 10^{12}$ & 0.787 &
$7.4 \times 10^{16}$ & $1.3 \times 10^{15}$ & 0.774 \\ 
Z4       & $1.0 \times 10^{15}$  & $6.7 \times 10^{12}$ & 0.807 & $7.8
\times 10^{16}$ & $1.0 \times 10^{15}$ & 0.779  \\  
Z8 (m=2) & $2.9 \times 10^{14}$  & $3.1 \times 10^{12}$ & 0.848 & $7.4
\times 10^{16}$ & $1.0 \times 10^{15}$ & 0.785 \\
\hline
         & \multicolumn{3}{c|}{Y4 (k=1, l=1)} & \multicolumn{3}{c|}{Y4
(k=2, l=0)} \\  
\hline
         & $\Lambda_Z$ & $\Lambda_Y$ & $g_{\rm string}$ & $\Lambda_Z$
& $\Lambda_Y$ & $g_{\rm string}$  \\
\hline
Z1       & $1.0 \times 10^{10}$   & $5.5 \times 10^{12}$ & 0.873 &  -
& - & - \\  
Z4       & $2.8 \times 10^{10}$  & $7.1 \times 10^{12}$ & 0.955 & 82 &
$1.2 \times 10^{13}$ & 1.376  \\  
Z8 (m=2) & $1.4 \times 10^{10}$   & $5.9 \times 10^{12}$ & 1.104 & - &
- & -\\ 
\hline
\end{tabular}
\end{center}
\caption{Same as Table~\ref{tbl:AAU}, only for case BYZ.}
\label{tbl:CCU}
\end{table}

\section{Models with Higher Kac-Moody Levels for \boldmath{$SU(2)_L$} or
\boldmath{$SU(3)_C$}}

We  assume that at the string
scale $M_{\rm string} \approx 5 \times 10^{17} ~{\rm GeV}$
the gauge couplings  satisfy
\begin{eqnarray}
g_1^2 = g_2^{\prime 2}= g_3^{\prime 2} ~,~\,
\end{eqnarray}
where
\begin{eqnarray}
g_1^2 \equiv k_Y g_Y^2~,~ g_2^{\prime 2}  \equiv k_2 g_2^{2}~,~
g_3^{\prime 2}  \equiv k_3 g_3^{2}~.~\,
\end{eqnarray}
Then, 
\begin{eqnarray}
b_1={{b_Y}\over {k_Y}}~,~b^{\prime}_2={{b_2}\over {k_2}}~,~
b^{\prime}_3={{b_3}\over {k_3}}~.~\,
\end{eqnarray}

It is very difficult to construct string models with
$k_2$ or $k_3$ larger than 2, and the discussions in models
 with $k_2=k_3=2$ are similar to those in Section IV by rescaling $k_Y$.
Therefore, we only study the models with $(k_2, k_3) = (1,2)$ or $(2,1)$.
The canonical $U(1)_Y$ normalization, $k_Y=5/3$,
 is not very interesting in these cases.  For brevity
we will not consider it here although the discussions would be similar
to those in Section IV.

\subsection{Models with $k_2=1$ and $k_3=2$}

To achieve string scale $SU(3)_C \times SU(2)_L$ unification, we need
to introduce sets of particles with $\Delta b_2 > \Delta b_3$.  With
the extra particles in Section II, we have the following 
independent Ti sets
\begin{eqnarray}
&& T1: \quad XQ + {\overline{XQ}} \,, \quad \Delta b =({1\over 5}, 3, 2)\,;\\ 
&& T2: \quad XL + {\overline{XL}} \,, \quad \Delta b = ({3\over 5}, 1, 0)\,;\\ 
&& T3: \quad XW \,, \quad \Delta b = (0, 2, 0)\,;\, \\
&& T4: \quad XW +  XU + {\overline{XU}} \,, \quad \Delta b = ({8\over 5}, 2, 1)\,;\\  
&& T5: \quad  XW + XD + {\overline{XD}} \,, \quad \Delta b = ({2\over 5}, 2, 1)\,;\\ 
&& T6: \quad 2 XW + XG \,, \quad \Delta b = (0, 4, 3)\,.\,
\end{eqnarray}

For simplicity, we only consider the cases with a single type of
particle set Ti but allow multiple copies.  In Table~\ref{tbl:k3e2}, we
list the numbers of the Ti sets necessary to ensure $SU(3)_C \times
SU(2)_L$ unification at the string scale, their mass scales, and the
corresponding non-canonical $U(1)_Y$ normalizations.  We can employ
the non-canonical normalizations as shown in the Table~\ref{tbl:k3e2}.

\begin{table}[htb]
\begin{center}
\begin{tabular}{|c|c|c|c|c|}
\hline
T's & $\Lambda_T$ & $k_Y$ & $k_Y$/(5/3) & $g_{\rm string}$ \\
\hline
$1 \times$ T1  & $6.1 \times 11^6$    & 58/15 & 2.320 & 1.227 \\ 
$2 \times$ T2  & $3.1 \times 10^4$    & 36/19 & 1.137 & 0.984 \\ 
$1 \times$ T3  & $5.2 \times 10^4$    & 8/3 & 1.600 & 0.984 \\
$2 \times$ T4  & $2.1 \times 10^9$    & 12/7 & 1.029 & 1.136\\
$2 \times$ T5  & $2.1 \times 10^9$    & 52/17 & 1.835 & 1.136 \\
$1 \times$ T6  & $7.2 \times 10^7$    & 29/6  & 2.900 & 1.344 \\
\hline
\end{tabular}
\end{center}
\caption{The mass scales and the corresponding $U(1)_Y$ normalizations
for Ti with $(k_2, k_3) = (1,2)$.}
\label{tbl:k3e2}
\end{table}

\subsection{Models with $k_2=2$ and $k_3=1$}

In this case we need to introduce Yi sets of particles.  The
numbers of the Yi sets, the mass scales at which they are
introduced and the appropriate $k_Y$ are shown in
Table~\ref{tbl:k2e2}.

\begin{table}[htb]
\begin{center}
\begin{tabular}{|c|c|c|c|c|}
\hline
Y's & $\Lambda_Y$ & $k_Y$ & $k_Y$/(5/3) & $g_{\rm string}$  \\ 
\hline
$3 \times$ Y2            & $8.0 \times 10^5$    & 11/5  & 1.320 &
1.032\\
\hline
$1 \times$ Y3            & 400                  & 61/10 & 3.660 &
1.511  \\
$1 \times$ Y3$^*$        & 2800                 & 25/4  & 3.750 &
1.561  \\
\hline
$1 \times$ Y4 (k=0, l=0) & $2.5 \times 10^{6}$  & 69/19 & 2.179 &
1.009  \\
\hline
$1 \times$ Y4 (k=0, l=1) & $3.5 \times 10^{4}$  & 61/19 & 1.926 &
1.169\\
$1 \times$ Y4$^*$ (k=0, l=1) & $2.0 \times 10^{5}$  & 59/18 & 1.967 &
1.203\\
\hline
$1 \times$ Y4 (k=0, l=2) & 450   & 43/11 & 2.345 & 1.492 \\
$1 \times$ Y4$^*$ (k=0, l=2) & 3100   & 4 & 2.400 & 1.540 \\
%$1 \times$ Y4 (k=1, l=0) & - & - & -\\ %$5.0 \times 10^6$    & 69/13 \\
%$1 \times$ Y4 (k=1, l=1) & - & - & -\\ %$2.7 \times 10^7$    & 65/7 & 2.041  \\
%$1 \times$ Y4 (k=2, l=0) & $1.5 \times 10^{10}$    & 250/13  & 2.784 \\
\hline
\end{tabular}
\end{center}
\caption{Same as Table~\ref{tbl:k3e2}, only for $(k_2, k_3) = (2,1)$.
The rows with an asterisk are for an effective SUSY scale of 50 GeV.}
\label{tbl:k2e2}
\end{table}

The Y3 and Y4 ($k=0,\ l=2$) cases imply small $\Lambda_Y$.  The values are
sensitive to the supersymmetric and string scale threshold
corrections, but the new particles may be observable at the LHC.  For the case with
Y4 ($k=0,\ l=0$), Y4 ($k=1,\ l=1$) and Y4 ($k=2,\ l=0$), we cannot achieve
string scale gauge coupling unification.

\section{Discussion and Conclusions}

Gauge coupling unification in the MSSM implies a unification scale
$M_U$ around $2\times 10^{16}$ GeV, while in weakly coupled
heterotic string theory the string scale $M_{\rm string}$ is about $ 5
\times 10^{17} ~{\rm GeV}$.  Because the latter is still one of the
leading candidates for a unified theory of the fundamental particles
and interactions, we studied the string scale gauge coupling
unification systematically by introducing vector-like particles whose
quantum numbers are the same as those of the SM fermions and their
Hermitian conjugates, and SM adjoint particles, and also allowed for
the possibility of non-canonical $U(1)_Y$ normalization.

We proposed a new approach.  We first considered the independent sets
Yi of particles that can be employed to achieve the $SU(3)_C$ and
$SU(2)_L$ string scale unification, and calculated the needed mass
scales.  We were then able to achieved string scale $SU(3)_C$,
$SU(2)_L$ and $U(1)_Y$ unification by choosing suitable $U(1)_Y$
normalizations for the Yi sets.  Alternatively, for canonical $U(1)_Y$
normalization, we considered suitable combinations of Yi sets
or introduced independent sets Zi with $\Delta b_2 = \Delta b_3$ in
addition to the Yi sets.  In a few cases the masses for all the extra
particles are roughly the same at the scale around $10^{15}$ GeV, or
else with one set around $10^{17}$ GeV, which can be considered as
string scale threshold corrections. In some cases there exist
additional particles with masses around hundreds of GeV with large
uncertainty from the supersymmetric and string thresholds, which may
be observable at the LHC.  Our approach can be
easily generalized to more complicated cases.  We also briefly
discussed string scale unification in models with higher Kac-Moody
levels for $SU(2)_L$ or $SU(3)_C$.

\section*{Acknowledgments}

 This research was supported in part by the U.S.~Department of Energy
 under Grants No.~DE-FG02-95ER40896 and DE-FG02-96ER40969, by the
 friends of the IAS, by the Cambridge-Mitchell Collaboration in
 Theoretical Cosmology, and by the University of Wisconsin Research
 Committee with funds granted by the Wisconsin Alumni Research
 Foundation.

\end{document}